\documentclass[aip, apl,reprint,longbibliography,superscriptaddress]{revtex4-1}
\usepackage{graphicx}
\usepackage{amsmath}
\usepackage{textcomp}
\usepackage[hidelinks]{hyperref}
\usepackage{xcolor}
\usepackage[utf8]{inputenc}
\graphicspath{{figures/}}

\graphicspath{{figures/}}

\begin{document}

\title{Measuring monopole and dipole polarizability of acoustic meta-atoms}

\author{Joshua Jordaan}
\affiliation{Nonlinear Physics Centre, Research School of Physics and Engineering, The Australian National University, Canberra, ACT 2601, Australia.}
\author{Stefan Punzet}
\affiliation{Nonlinear Physics Centre, Research School of Physics and Engineering, The Australian National University, Canberra, ACT 2601, Australia.}
\affiliation{Faculty of Electrical Engineering and Information Technology, Ostbayerische Technische Hochschule Regensburg, Seybothstraße 2, 93053 Regensburg, Germany}
\affiliation{Department of Electrical and Computer Engineering, Technical University of Munich, Theresienstr.~90, 80333 Munich, Germany}
\author{Anton Melnikov}
\affiliation{Chair of Vibroacoustics of Vehicles and Machines, Technical University of Munich, Boltzmann Str.~15, 85748 Garching, Germany}
\affiliation{Centre for Audio, Acoustics and Vibration, University of Technology Sydney, NSW 2007, Australia}
\affiliation{School of Engineering and Information Technology, University of New South Wales, Canberra, ACT 2610, Australia.}
\author{Alexandre Sanches}
\affiliation{Nonlinear Physics Centre, Research School of Physics and Engineering, The Australian National University, Canberra, ACT 2601, Australia.}
\affiliation{School of Engineering, University of São Paulo, Av. Prof. Luciano Gualberto, 380 - Butantã, CEP 05508-010, São Paulo, SP, Brazil}
\author{Sebastian Oberst}
\affiliation{Centre for Audio, Acoustics and Vibration, University of Technology Sydney, NSW 2007, Australia}
\author{Steffen Marburg}
\affiliation{Chair of Vibroacoustics of Vehicles and Machines, Technical University of Munich, Boltzmann Str.~15, 85748 Garching, Germany}
\author{David A.~Powell}
\email{david.powell@adfa.edu.au}
\affiliation{School of Engineering and Information Technology, University of New South Wales, Canberra, ACT 2610, Australia.}
\affiliation{Nonlinear Physics Centre, Research School of Physics and Engineering, The Australian National University, Canberra, ACT 2601, Australia.}

\begin{abstract}

We present a method to extract monopole and dipole polarizability from experimental measurements of two-dimensional acoustic meta-atoms. In contrast to extraction from numerical results, this enables all second-order effects and uncertainties in material properties to be accounted for. We apply the technique to 3D-printed labyrinthine meta-atoms of a variety of geometries. We show that the polarizability of structures with shorter acoustic path length agrees well with numerical results. However, those with longer path lengths suffer strong additional damping, which we attribute to the strong viscous and thermal losses in narrow channels.

\end{abstract}

\maketitle

Acoustic metasurfaces are metamaterial structures with sub-wavelength thickness that can implement a rich variety of acoustic functions \cite{zhao_manipulation_2016,CummerControllingsoundacoustic2016}. A promising approach for metasurfaces is the design of structures with internal labyrinthine configuration to slow down the acoustic wave's velocity to create compact resonators \cite{liang_extreme_2012,XieTaperedlabyrinthineacoustic2013}. Structures of this kind exhibit excellent wavefront shaping potential \cite{li_experimental_2014,XieWavefrontmodulationsubwavelength2014,shen_design_2016,zhao_manipulation_2016}.
Such meta-atoms can generate phase shifts up to $2\pi$ by adjusting their geometry \cite{shen_design_2016}. Thereby, a wave manipulation function can be realized with a corresponding phase gradient, which is then discretized to enable implementation with an array of meta-atoms.  

Drawing inspiration from electromagnetism, the dominant design paradigm for acoustic metasurfaces has been the generalized Snell's law \cite{li_experimental_2014,tang_anomalous_2014}, where structures are designed for high amplitude, with spatially varying phase, both for transmission or reflection problems. However, in electromagnetism, it has been shown that the generalized Snell's law does not correctly account for impedance matching and energy conservation.  Approaches based on surface impedance must be used instead \cite{achouri_general_2015,estakhri_wave-front_2016} and equivalent electric and magnetic surface impedances need to be defined. Recently these more accurate surface-impedance models have also been applied to acoustic metasurfaces \cite{Diaz-RubioAcousticmetasurfacesscatteringfree2017,LiSystematicdesignexperimental2018}. The impedances can be derived from the multipole moments of a single meta-atom \cite{kuester_averaged_2003}. In the acoustics of fluids, the fundamental moments are the monopole and dipole, corresponding to the net compression and displacement of a fluid volume respectively. The acoustic response of sub-wavelength meta-atoms is well-approximated by their monopole and dipole polarizability coefficients. These coefficients relate the strength of the monopole and dipole moments to the incident pressure and velocity fields respectively. Developing a model based on polarizability can lead to great simplifications in modelling, particularly for complex arrangements of meta-atoms.

An alternative to  a continuously connected metasurface is the use of sparse arrays of disconnected resonant meta-atoms\cite{zhao_manipulation_2016,cheng_ultra-sparse_2015}, which can enable highly efficient beam refraction at large angles \cite{QuanMaximumWillisCoupling2018}.  These elements may find their application in creating sound control structures which also allow airflow. Here, the monopole and dipole polarizabilities of the meta-atoms are the most natural model to apply. To date these polarizabilities have not been directly measured; with most designs relying on simulations or indirect observations of resonances attributed to the monopolar and dipolar modes \cite{cheng_ultra-sparse_2015,KrushynskaSpiderwebstructuredlabyrinthine2017}. 

In this work we present a technique for directly extracting the acoustic monopole and dipole polarizability of two-dimensional meta-atoms from experimental measurements. In addition, the method can be applied to numerically extracted data. Obtaining polarizability information from experimental measurements is necessary for good accuracy, since numerical simulation may neglect  viscous and thermal boundary layers, the excitation of vibration modes in thin structures, and it may be difficult to obtain reliable material properties for rapid prototyping materials.

\begin{figure}[tb]
	\centering
    	\includegraphics[width=\columnwidth]{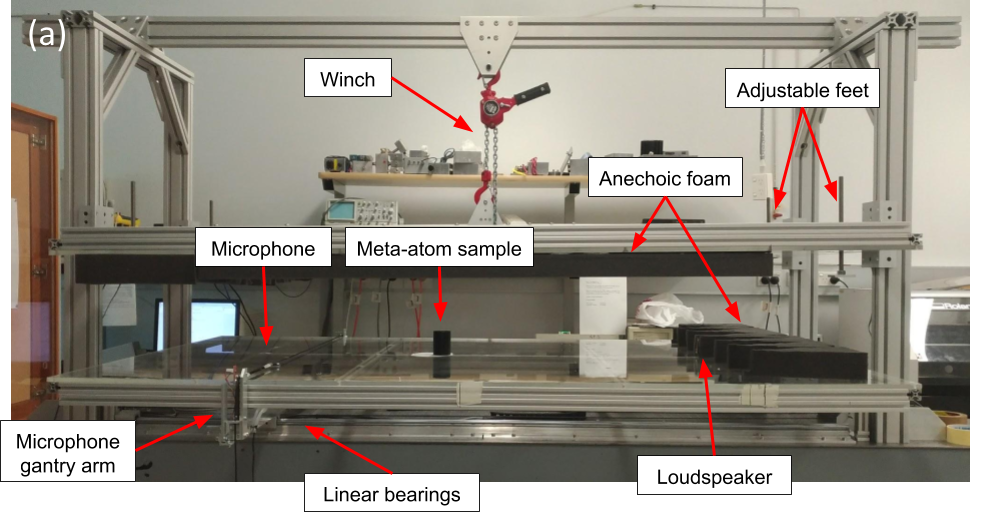}
		\includegraphics[width=\columnwidth]{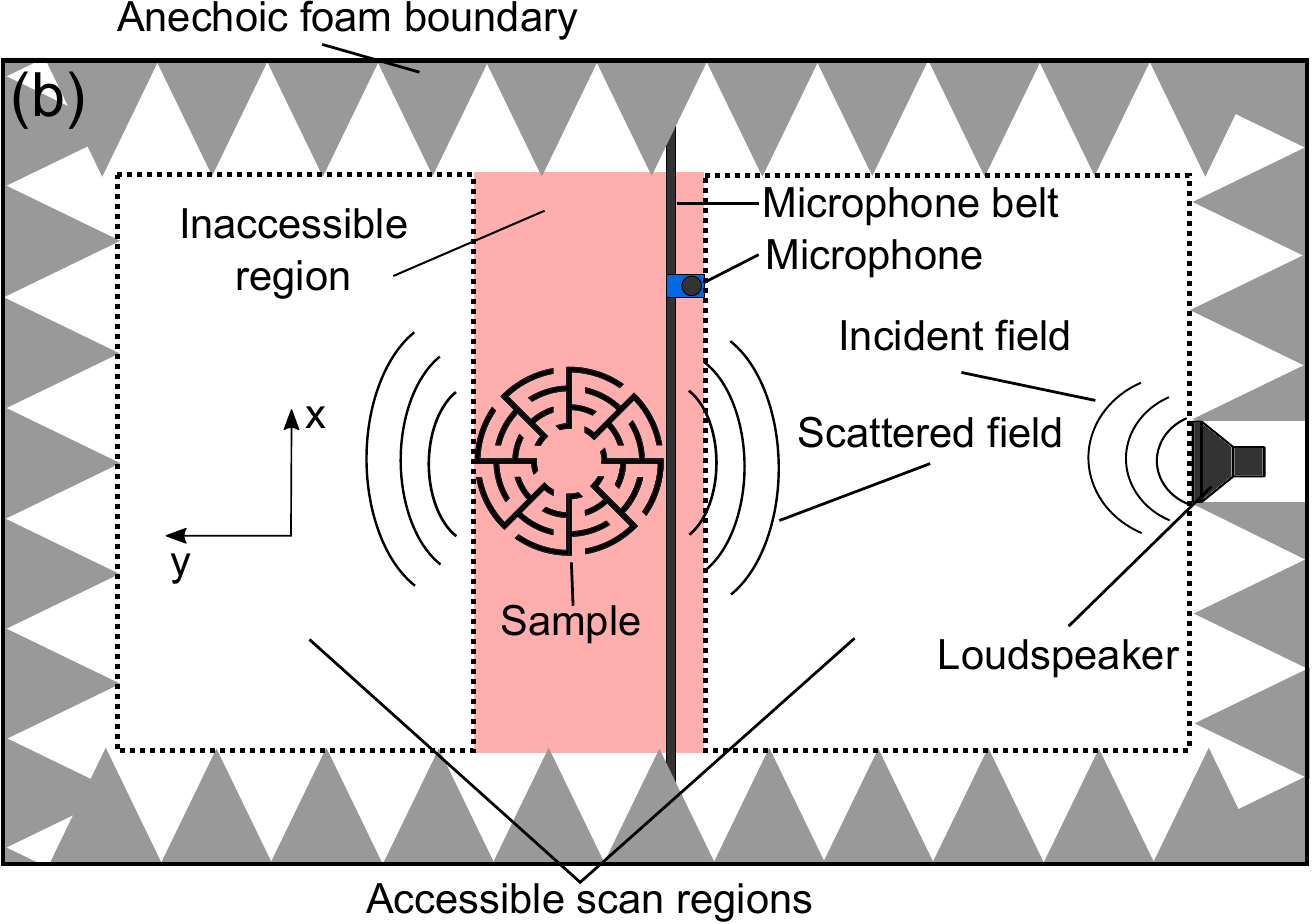}
	\caption{(a) Side-view of the parallel plate waveguide apparatus, where the lid has been raised to access the sample. (b) Schematic diagram of the  apparatus and sample (not to scale), showing a top-down view of components within the waveguide.}
	\label{fig:schematic}
\end{figure}

In this work we consider the experimental configuration shown in Fig.~\ref{fig:schematic}, similar to that used in previous works \cite{ZigoneanuDesignmeasurementsbroadband2011a,li_experimental_2014}. Two plates separated by a 66\,mm gap form a parallel-plate acoustic waveguide, with uniform pressure distribution in the vertical direction, making it effectively a two-dimensional (2D) system. A loudspeaker at one end acts as the acoustic source while foam wedges form an absorbing boundary.  A microphone mounted on a belt system can be scanned to any position within the $x-y$ plane of the 2D acoustic system. A micro-controller development board with audio peripheral (Teensy 3.2 with Audio Adapter Board\cite{teensy_audio}) digitally generates and coherently detects the sinusoidal waves. Using the internally generated source as a phase reference eliminates the need for a two microphone measurement. The amplitude and phase response of the speakers, microphone and amplifier are unknown, but are eliminated in the extraction procedure outlined below, by explicitly measuring the incident field as a reference. Details of the construction and initial characterization of this system can be found in Refs.~\onlinecite{JordaanAcousticMetaatomsExperimental2017,PunzetDesignconstructionAcoustic2016}.

For the corresponding numerical analysis, we use a custom 2D boundary element method (BEM) code.
This treats all solids as acoustic hard boundaries, ignores losses in air, and solves for the scattered pressure on the surface of the object.
The code uses continuous elements with quadratic interpolation functions and discretization
by collocation method with an adaptive integration scheme \cite{marburg_boundary_2018}.

The experimental apparatus shown in Fig.~\ref{fig:schematic} essentially yields the scalar pressure field throughout a plane. The parallel plate waveguide is operated in a regime where it supports only the fundamental mode with uniform pressure along the $z$ direction. The first higher order modewith inhomogeneous pressure distribution in the $z$ direction\cite{kinsler_fundamentals_2000} becomes propagating at 2,598\,Hz, setting the upper limit of measurement. 
The pressure field $p_\mathrm{inc}$ of an arbitrary incident wave propagating within a 2D system can be expanded as
\begin{equation} \label{eq:bessel_inc}
p_\mathrm{inc}(r, \theta) = \sum_{n = -\infty}^{\infty}\beta_nJ_n(k_0r)e^{in\theta},
\end{equation}
where $J_n$ is the Bessel function of the first kind, $k_0 = \omega/c$ is the propagation wavenumber, $\omega$ and $c$ are the incident wave's natural frequency and the speed of sound in air,  $(r,\theta)$ are the polar coordinates relative to the center of the sample, $i$ is the imaginary unit and  $\beta_n$ are the expansion coefficients. These expansion coefficients can  be found analytically only for specific incident field profiles such as a plane wave \cite{mei_mathematical_1997}.

In our apparatus the incident field generated by the single speaker shows significant curvature of the wave-fronts, and a spatially inhomogeneous amplitude, thus it is not well approximated by a plane wave. Furthermore the source amplitude varies strongly with frequency due to the speaker response, requiring that $\beta_n$ are fitted to the measured acoustic field.  For the calculation of polarizability, both the incident pressure and the velocity at the center of the scatterer can be directly retrieved from the expansion coefficients as $p_\mathrm{inc}(0) = \beta_0$ and $v_{\mathrm{inc},y}(0) = -\frac{\beta_1+\beta_{-1}}{2c\rho_0}$, with $\rho_0$ being the density of air.

The corresponding scattered pressure field $p_\mathrm{scat}$ can be expanded as
\begin{equation}\label{eq:hankel_scat}
p_\mathrm{scat}(r, \theta) =  \sum_{n = -\infty}^{\infty}\gamma_nH_n^{(1)}(k_0r)e^{in\theta},
\end{equation}
where $H_n^{(1)}$ is the Hankel function of the first kind. The expansion coefficients $\gamma_n$ can be related to the dominant monopole and dipole moments as $M=\frac{4i}{\omega^2}\gamma_0$ and $D_y=\frac{4c}{\omega^3}(\gamma_1+\gamma_{-1})$.
For a scatterer of arbitrary shape, the scattering process is described by a full tensor $\gamma_n =\sum_m S_{nm}\beta_n$. However, for objects with approximate circular symmetry, all terms $n\neq m$ are zero, and the relationship becomes scalar: $\gamma_n=S_{nn}\beta_n$ with $S_{-n-n}=S_{nn}$. 

For sub-wavelength meta-atoms, the monopolar and dipolar terms are expected to dominate scattering. For analytical modelling of collections of meta-atoms it is more convenient to use monopole and dipole polarizability coefficients $\alpha_{pp}$ and $\alpha_{vv}$ satisfying
\begin{equation}
M=\alpha_{pp}p_\mathrm{inc}(0), \qquad \textbf{D} = \alpha_{vv}\mathbf{v}_\mathrm{inc}(0),
\end{equation}
where $\alpha_{vv}$ becomes a scalar for the rotationally symmetric structures considered here. However, by normalizing these polarizabilities, we find that they are trivially related to the scattering coefficients:
\begin{align}
\alpha_{pp}' &= \frac{\omega^2}{4i}\alpha_{pp} = S_{00}  \label{eq:monopole_normalized}\\
\alpha_{vv}' &= \frac{-\omega^3}{8c^2\rho_0}\alpha_{vv} = S_{11} = S_{-1-1}  \label{eq:dipole_normalized}
\end{align}
We use this normalization since it gives a simple physical interpretation of the strength of different types of polarizability in terms of contribution to scattering, with a maximum magnitude of unity.

To experimentally measure the acoustic polarizability, we need to determine the incident field coefficients $\beta_n$ and the scattered field coefficients $\gamma_n$ for $n \in \{0,1,-1\}$. The incident field is measured on a circle of radius $R_\mathrm{inc}$. Applying the orthogonality of exponential functions to Eq.~\eqref{eq:bessel_inc}, we find the incident field coefficients as:
\begin{equation} \label{eq:incident_coefficients}
\beta_n = \frac{1}{2\pi J_n(k_0R_\mathrm{inc})}\int_{-\pi}^\pi p_\mathrm{inc}(R_\mathrm{inc},\theta)e^{-in\theta}\mathrm{d}\theta.
\end{equation}
Note that the Bessel functions have zeros which make Eq.~\eqref{eq:incident_coefficients} singular, the first of which occurs at $k_0R_\mathrm{inc}\approx2.4$. Thus $R_\mathrm{inc}$ is chosen sufficiently small to remain well below this singular condition at the highest frequency of interest.

For determination of the scattered field coefficients, we measure both the total field $p_\mathrm{tot}$ and incident field $p_\mathrm{inc}$ at the same radius $R_\mathrm{scat}$, with the scattered field given by their difference $p_\mathrm{scat}=p_\mathrm{tot}-p_\mathrm{inc}$. Integrating Eq.~\eqref{eq:hankel_scat} and applying orthogonality conditions, the scattered field coefficients are given by
\begin{equation} \label{eq:scattered_coefficients}
\gamma_n = \frac{1}{2\pi H^{(1)}_n(k_0R_\mathrm{scat})}\int_{-\pi}^\pi p_\mathrm{scat}(R_\mathrm{scat},\theta)e^{-in\theta}\mathrm{d}\theta.
\end{equation}
As the Hankel function has no real zeros, there is more freedom to choose $R_\mathrm{scat}$. Referring to Fig.~\ref{fig:schematic}(b), we see that when the sample is placed within the waveguide, there is an inaccessible region of width $w$ where the field cannot be measured, as the belt on which the microphone is mounted would collide with the sample. Therefore we must measure over a reduced angular range, and approximate the angular integral as
\begin{equation}
\tfrac{1}{2\pi}\int_{0}^{2\pi}\cdots\mathrm{d}\theta\approx\tfrac{1}{2\pi-4\theta_{in}}\left(
\int_{\theta_{in}}^{\pi-\theta_{in}}\cdots\mathrm{d}\theta +
\int_{\pi+\theta_{in}}^{2\pi-\theta_{in}}\cdots\mathrm{d}\theta
\right),
\end{equation}
where $\theta_{in}=\arcsin\frac{w}{2R_\mathrm{scat}}$ is the angular half-width of the inaccessible region. Since the range of inaccessible angles reduces with increasing $R_\mathrm{scat}$, larger values are preferred for increased accuracy. This reduced angular range of integration means that we do not have exact orthogonality between different orders $n$. However, for sub-wavelength meta-atoms with dominant monopolar and dipolar radiation the scattered field will have relatively smooth angular variation, and we do not expect significant interference from higher order terms with $|n| > 1$.

\begin{figure}[tb]
	\centering
	\includegraphics[width=\columnwidth]{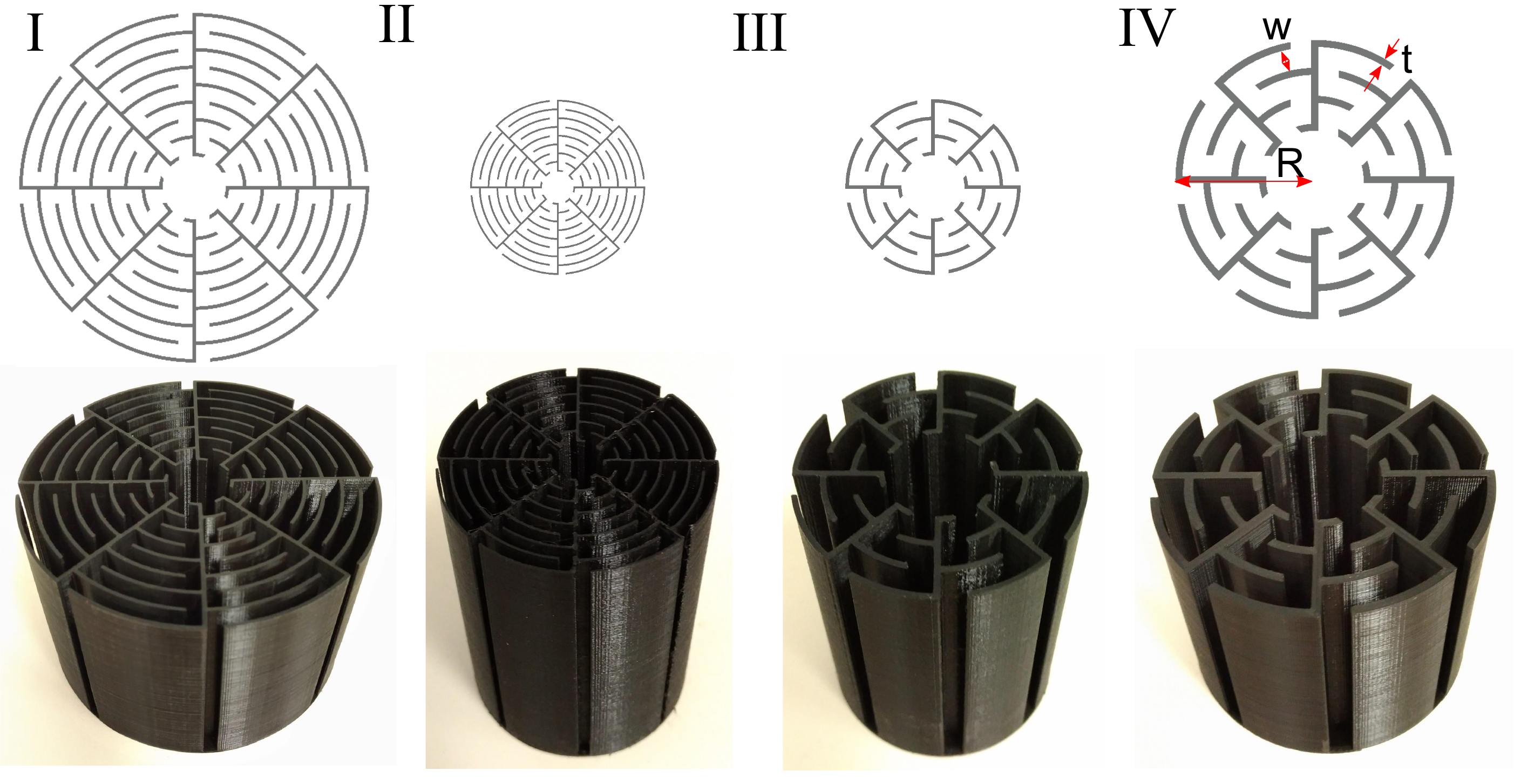}
	\caption{Top row: Cross-section of each meta-atom design. Meta-atom I was taken from Ref.~\onlinecite{cheng_ultra-sparse_2015}, IV from Ref.~\onlinecite{lu_realization_2017}. The diagram of meta-atom IV defines the parameters wall thickness $t$, channel width $w$ and meta-atom radius $R$. For I: $R = 50$\,mm, $w = 4$\,mm and $t = 1$\,mm. II: $R = 25$\,mm, $w = 2$\,mm, $t = 0.5$\,mm. III: $R = 25$\,mm, $w = 4$\,mm, $t = 1$\,mm. IV: $R = 40$\,mm, $w = 6$\,mm, t = 2 mm. Bottom row: Photographs of the 3D printed meta-atoms, made from PLA with 0.1\,mm layer thickness to a height of 66\,mm.}
	\label{fig:fabricated_structures}
\end{figure}

The ratio of the scattered field coefficients $\gamma_n$ to the incident field coefficients $\beta_n$ gives the corresponding scattering coefficient $S_{nn}$ according to Eqs.~\eqref{eq:monopole_normalized} and \eqref{eq:dipole_normalized}. Since we have two equivalent expressions for the dipole polarizability $\alpha_{vv}'$, their average is taken to reduce the influence of measurement uncertainties.

The developed extraction procedure is applied to individual 2D acoustic meta-atoms based on labyrinthine designs with eight-fold rotational symmetry. Four meta-atoms were fabricated and characterized, two of them having geometries previously reported in Refs.~\onlinecite{cheng_ultra-sparse_2015,lu_realization_2017}. Diagrams of the designs are shown in Fig.~\ref{fig:fabricated_structures}, with photographs of the fabricated meta-atoms shown below. All of the designs were fabricated by 3D printing using PLA filament with a 0.1\,mm layer thickness, to a height of 66\,mm. The top of each meta-atom was left open to simplify fabrication, and to allow verification of the fabrication quality. Initial experiments with this configuration showed poor agreement with the numerical results, due to imperfect contact between the meta-atom and the top waveguide plate. We attributed these poor initial results to sound leakage from a small gap between the meta-atom and top plate, and the excitation of vibrational modes in the meta-atoms. To suppress these effects we inserted a thin rubber sheet between the meta-atom and top plate, which greatly improved agreement, as detailed below.

\begin{figure}[tb]
	\centering
	\includegraphics[width=\columnwidth]{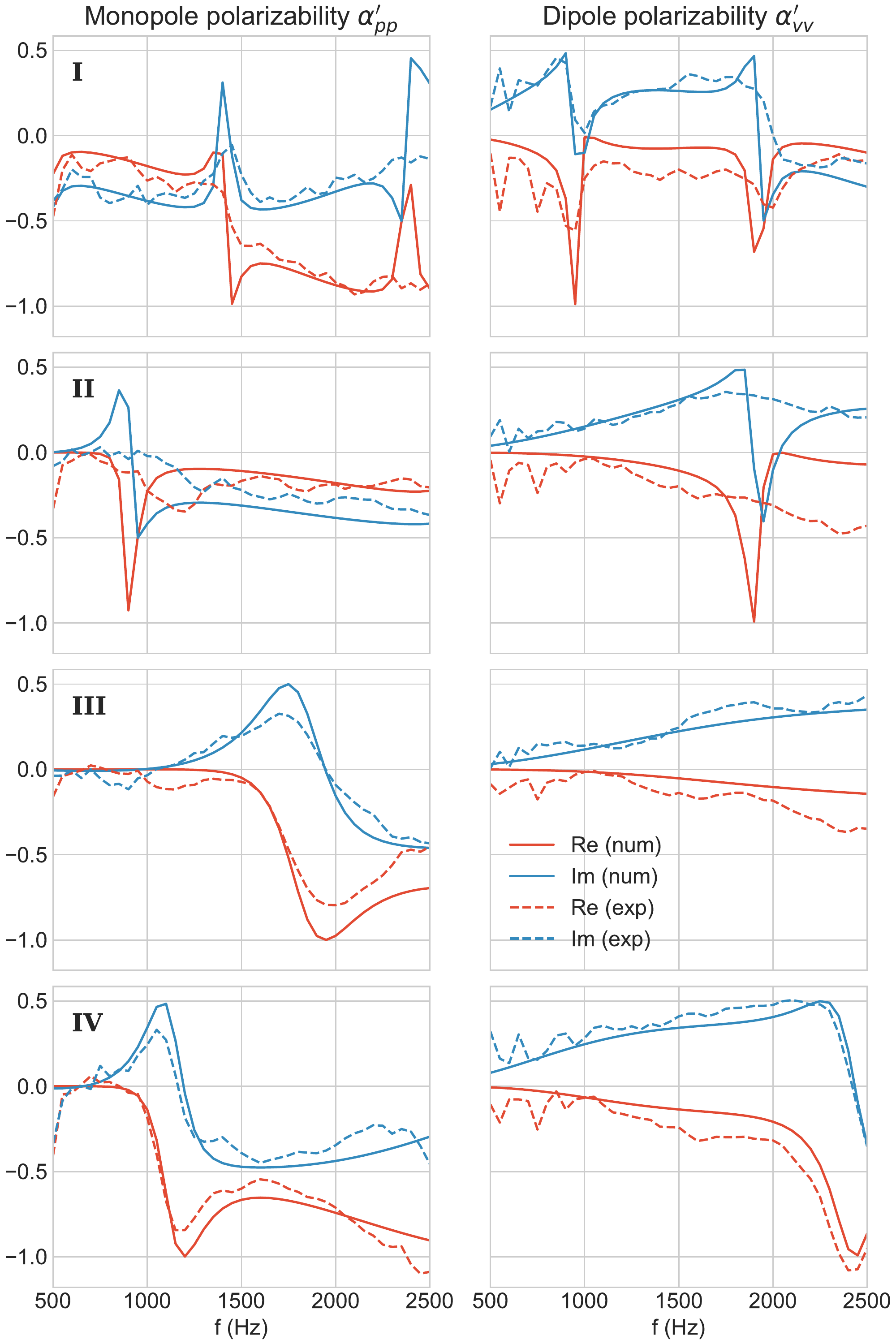}
	\caption{Experimentally measured (dashed lines) and numerically calculated (solid lines) normalized monopole and dipole polarizability.}
	\label{fig:metaatom_polarizability}
\end{figure}

The experimentally measured (dashed curves) and numerically calculated (solid curves) polarizabilities are shown  in Fig.~\ref{fig:metaatom_polarizability}. Note that we consider only frequencies above 500\,Hz, since data at lower frequencies are inaccurate due to the poor performance of the absorbing boundaries at long wavelengths, as well as high background noise levels in that frequency range. Overall the agreement is reasonable, but the experimental results contain small spurious peaks and ripples, and in certain cases resonant features predicted in the numerical results are heavily suppressed. A possible explanation for the ripples are multiple reflections between the meta-atom and the speaker. These multiple reflections result in an effective incident field that differs from that field, measured in the absence of the meta-atom. On the other hand, the suppression of certain resonances is a genuine physical effect, as discussed below.

Consider first the numerical results for meta-atom I, corresponding to the geometry reported in Ref.~\onlinecite{cheng_ultra-sparse_2015}. In that work, the authors performed a semi-analytical derivation of the multipole expansion coefficients of the field scattered by a plane-wave (comparable to $S_{nn}$). They predicted a fundamental monopole resonance at 518\,Hz and a dipole resonance at 1,080\,Hz, as well as a second monopole resonance at 1,549\,Hz. We see that our numerical results are fairly consistent with these semi-analytical predictions, however the numerically determined resonant frequencies are somewhat lower. This has the effect of pushing the fundamental monopole resonance below our measurement range. To enable us to characterize this resonance, we designed meta-atom II, where all geometric parameters are downscaled by 1/2. The numerically predicted fundamental dipole and monopole resonances are now moved to double the frequency, and the fundamental monopole resonance is within the measurable range.

Comparing the experimental results of meta-atoms I and II to their numerical results, indicates the presence of resonant peaks, which are suppressed to the level of the experimental uncertainties. This is most likely due to the viscous and thermal boundary layers, which are not included in our boundary element model. These layers have been shown to dominate the acoustic response in channels\cite{WardBoundarylayereffectsacoustic2015}, lowering the quality factor of resonances. 
To reduce the influence of losses, we created meta-atom III. This had the same channel thickness $w$ and wall thickness $t$ as meta-atom I, but with the external radius being scaled by 1/2. Due to the greatly reduced acoustic path-length, the fundamental monopolar resonance is thereby increased to approximately 1,900\,Hz, and the dipolar resonance is outside the measurable frequency range. For this structure, it can be seen that the numerical results agree much better with the experiments. 

To further investigate the applicability of our extraction method, we also fabricated meta-atom IV, which was originally presented in Ref.~\onlinecite{lu_realization_2017} with a predicted fundamental monopole resonance at 1,360\,Hz. Meta-atom IV has thicker walls and wider channels than the other meta-atoms; thus it is expected that this structure is less susceptible to boundary layer effects. It can be seen that the experimental and numerical results are in good agreement for this structure. The resonant frequency is lower than predicted in Ref.~\onlinecite{lu_realization_2017}, being approximately 1,200\,Hz in both numerics and experiment. The good agreement between numerical and experimental results for this sample further validates our approach.

To confirm that the viscous and thermal boundary layers are responsible for suppressing the scattering peaks of meta-atoms I and II, we apply the theory of Stinson \cite{Stinsonpropagationplanesound1991} to calculate the complex wavenumber $k_\mathrm{eff}$ within the narrow channels. We consider one of the channels in each meta-atom, and approximate it as a straight rectangular pipe with length $l_\mathrm{eff}$ given by the acoustic path length from the center cavity to the exterior. The damping within the pipe yields its dissipative quality factor as \cite{kinsler_fundamentals_2000}
\begin{equation}
Q_\mathrm{diss} = \frac{\mathrm{Re}\{k_\mathrm{eff}\}}{2\mathrm{Im}\{k_\mathrm{eff}\}}.
\end{equation}
We also take into account the leakage of energy through the external aperture by each channel. This can be calculated from the radiation resistance of an unflanged pipe \cite{kinsler_fundamentals_2000}, to yield the radiative quality factor
\begin{equation}
Q_\mathrm{rad} = \frac{4\pi l_\mathrm{eff}}{S\mathrm{Re}\{k_\mathrm{eff}\}},
\end{equation}
where $S$ is the cross-sectional area of the aperture. These quality factors were calculated for the lowest order monopole mode of each meta-atom within our experimental frequency range, and are listed in Table \ref{tab:q_factors}.

\begin{table}
\caption{\label{tab:q_factors}Estimated dissipative and radiative quality factors of the lowest frequency monople resonances shown in Fig.~\ref{fig:metaatom_polarizability}.}
\begin{ruledtabular}
\begin{tabular} {r c c c c}
Structure & I & II & III & IV \\
$Q_\mathrm{diss}$  & 36  & 12  & 43 & 43\\
$Q_\mathrm{rad}$ & 146 & 142 & 37 & 39\\
\end{tabular}
\end{ruledtabular}
\end{table}

The dissipative and radiative quality factors can be combined to find the total quality factor $Q_\mathrm{tot} = \left(Q_\mathrm{diss}^{-1} + Q_\mathrm{rad}^{-1}\right)^{-1}$, which determines the total rate of energy loss from the cavity. However, to interpret the scattering response, we need to consider the individual contribution from each of these terms. By reciprocity, the radiative quality factor $Q_\mathrm{rad}$ also determines how long it takes an incident wave to couple into the structure. $Q_\mathrm{diss}$ determines the time scale over which energy is dissipated internally. If $Q_\mathrm{diss} \ll Q_\mathrm{rad}$, then internal dissipation will dominate over radiation of energy, and resonant scattering will be suppressed. As can be seen in Table \ref{tab:q_factors}, for meta-atoms I and II, internal dissipation dominates over radiation, which explains the strong suppression of the resonant scattering peaks. In contrast, for meta-atoms III and IV, radiative and dissipative losses are comparable, thus the experimentally observed scattering is only moderately suppressed compared to the simulated values.

In conclusion, we presented a method for extracting the monopole and dipole polarizability from experimental measurements of two-dimensional acoustic meta-atoms. We applied this method to labyrinthine meta-atoms previously reported in the literature. For structures with thin walls and long acoustic path length, the resonances predicted numerically were highly damped, and were essentially unobservable in the experiment. We attribute this to the viscous and thermal boundary layers, which have thickness comparable to the width of the narrow channels. When applying our method to structures with shorter acoustic path lengths and wider channels, we found good agreement with numerical results.

We acknowledge useful discussions with Andrea Al\`u and Li Quan. AM acknowledges the financial support provided by SO over the UTS Centre for Audio, Acoustics and Vibration (CAAV) international visitor funds. DP acknowledges funding from the Australian Research Council through Discovery Project DP150103611.

\end{document}